\documentclass{mem}
\usepackage{natbib}\usepackage{txfonts}\usepackage{balance}
\usepackage{graphicx}
\usepackage[a4paper,breaklinks,dvipdfm]{hyperref}
\idline{00}{000}
\begin{document}
\def\teff{$T\rm_{eff }$}
\def\kms{$\mathrm {km s}^{-1}$}

\title{Non-thermal states in models of filaments: a dynamical study}
\subtitle{}

\author{P. \,Di Cintio\inst{1,2}\and S. \,Gupta\inst{3}\and L. \,Casetti\inst{2,4,5}}

\institute{Consiglio Nazionale delle Ricerche, Istituto di Fisica Applicata ``Nello Carrara", via Madonna del piano 10, I-50019 Sesto Fiorentino, Italy
\and INFN -  Sezione di Firenze, via G.\ Sansone 1, I-50019 Sesto Fiorentino, Italy 
\and Department of Physics, Ramakrishna Mission Vivekananda University, Belur Math, Dist Howrah 711202 West Bengal, India
\and Dipartimento di Fisica e Astronomia and CSDC, Universit\`a di Firenze, via G.\ Sansone 1, I-50019 Sesto Fiorentino, Italy
\and INAF - Osservatorio Astrofisico di Arcetri, largo Enrico Fermi 5, I-50125 Firenze, Italy
\email{p.dicintio@ifac.cnr.it}}
\authorrunning{Di Cintio, Gupta \& Casetti}
\titlerunning{Non-thermal states in filaments}
\abstract{
We study the origin of the non-thermal profiles observed in filamentary structures in galactic molecular clouds by means of numerical dynamical simulations. We find that such profiles are intrinsic features of the end products of dissipationless collapse in cylindrical symmetry. Moreover, for sufficiently cold initial conditions, we obtain end states characterized by markedly anticorrelated radial density and temperature profiles. Gravitational, dissipationless dynamics alone is thus sufficient to reproduce, at least qualitatively, many of the properties of the observed non-thermal structures.  
}
\maketitle{}
Filaments in galactic molecular clouds are (at least in their initial stages) mainly gravitationally supported structures, that also harbor star-forming cores (see e.g.\ \citealt{fede}). 
Remarkably, observations suggest that filaments are in non-thermal states \citealt{arz}; a good description seems to be given by polytropic equations of state $\rho \propto T^n$ or $P \propto \rho^\gamma$ (\citealt{toci1,toci2}). In general it is believed that this is due to the interplay between local turbulence, stellar feedback, radiation transport, and magnetic fields.

As a very simple model of a filament we have considered an infinite self-gravitating cylinder, whose dynamics is thus mapped onto that of a two-dimensional system of self-gravitating particles with logarithmic interactions \citep{pf3}, neglecting all contributions arising from magnetic fields and radiation. We have performed numerical simulations of the dynamics by means of direct $N$-body integration as well as of 2D particle-in-cell (PIC), also including multiparticle collisions (MPC) (see \citealt{pf1,pf2}).  
\begin{figure*}[]
\resizebox{0.9\hsize}{!}{\includegraphics[clip=true]{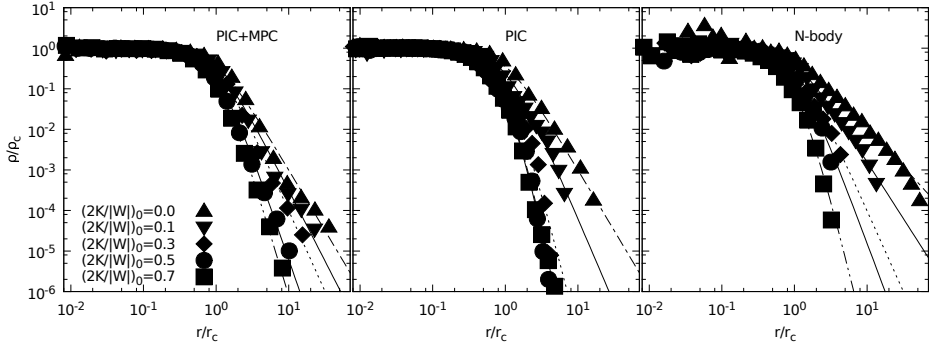}}
\caption{
\footnotesize
Radial density profile of the end states of collapsing Gaussian overdensities with different values of the initial virial ratio (points) and best fit curves (lines) for (from left to right) PIC+MPC, PIC and $N-$body simulation protocols.
}
\label{endstates}
\end{figure*}
We follow the collapse of a cold and gravitationally unstable (i.e., with initial virial ratio $2K/|W|<1$) cylindrically symmetric overdensity with Gaussian radial density profile. After a violent contraction phase, the system relaxes to a structure with radial density profile fitted (see Fig.\ \ref{endstates}) by $\rho(r)={\rho_cr_c^\alpha}{(r_c^2+r^2)^{-\alpha/2}},$
where $\rho_c$ and $r_c$ are the core density and core radius, respectively. The latter nicely approximates the density profile of a polytropic filament.
For sufficiently low initial kinetic temperatures (corresponding to $2K/|W|\leqslant 0.1$), we find values of $\alpha$ of the order of 2 \citep{pf3}, rather close to the observed systems (see \citealt{arz}). Moreover the kinetic temperature profiles of such systems have strongly increasing gradients for increasing $r$. Such feature is also observed in real filaments, and in the end-products of simulations where an isothermal \citep{ostriker} filament suffers a strong radial perturbation. Moreover, anticorrelated density vs.\ temperature profiles have been also found in non-astrophysical contexts, for instance in mean-field models kicked out of equilibrium by an impulsive perturbation (\citealt{casetti,teles1}) or in condensed matter systems (\citealt{gupta}). As a possible mechanism to explain why these non-thermal states exhibit temperature inversion \cite {teles1} suggested that during the initial violent relaxation phase the interaction of the particles with the collective oscillations may produce suprathermal tails in the velocity distribution function. In an inhomogeneous system, this may trigger a ``velocity filtration'' mechanism (\citealt{scudder1}) broadening the velocity distribution function where the system is less dense, because only sufficiently fast particles may escape the potential well produced by the central concentration.

In conclusion, it appears that dissipationless collapse alone can produce dynamically supported non-thermal end states qualitatively similar to those observed in filaments. Some instances exhibit marked anticorrelated temperature and density profiles, and non-thermal long-lived states with these features may occur in any long-range-interacting system after the damping of collective oscillations. 
\bibliographystyle{aa}

\begin{thebibliography}{}
\bibitem[Arzoumanian et al. (2011)]{arz} Arzoumanian, D., et al.\ 2011, A\&A 529, 6 
\bibitem[Casetti \& Gupta (2014)]{casetti} Casetti, L., \& Gupta, S.\ 2014,  EPJB 87, 91
\bibitem[Di Cintio et al. (2015)]{pf1} Di Cintio, P., Livi, R., Bufferand, H., Ciraolo, G., Lepri, S., \& Straka, M.J.\ 2015, Phys. Rev. E 92, 062108
\bibitem[Di Cintio et al. (2017)]{pf2} Di Cintio, P., Livi, R. Lepri, S. \& Ciraolo, G.\ 2017, Phys. Rev. E 95, 043203
\bibitem[Di Cintio, Gupta \& Casetti (2017)]{pf3} Di Cintio, P., Gupta, S., \& Casetti, L.\ 2017, arXiv:1706.01955
\bibitem[Federrath et al.\ (2016)]{fede} Federrath, C., et al.\ 2016, ApJ 832, 143
\bibitem[Gupta \& Casetti (2016)]{gupta} Gupta, S., \& Casetti L.\ 2016, New Journal of Physics 18, 103051 
\bibitem[Ostriker (1964)]{ostriker} Ostriker, J.\ P.\ 1964, ApJ 140, 1056 
\bibitem[Scudder (1992a)]{scudder1} Scudder, J.\ 1992, ApJ 398, 299 
\bibitem[Teles et al.\ (2015)]{teles1} Teles, T.\ N., Gupta, S., Di Cintio, P., \& Casetti, L.\ 2015, Phys. Rev. E 92, 020101 
\bibitem[Toci \& Galli (2014a)]{toci1} Toci, C., \&  Galli, D.\ 2014, MNRAS 446, 2110 
\bibitem[Toci \& Galli (2014b)]{toci2} Toci, C., \&  Galli, D.\ 2014, MNRAS 446, 2118 
\end{thebibliography}

\end{document}